\newcommand{\beq}{\begin{eqnarray}}
\newcommand{\eeq}{\end{eqnarray}}

\newcommand{\dota}{\dot{a}}

\documentclass[prb]{revtex4}
\usepackage{amsmath}
\usepackage{graphicx}

\begin{document}

\title{A physical interpretation of Hubble's law and the cosmological redshift from the perspective of a static observer}

\author{Simen Braeck}
\email{Simen.Brack@hioa.no}
\affiliation{Oslo and Akershus University College of Applied Sciences, Faculty of Engineering, P.O. Box 4 St. Olavs Plass, N-0130 Oslo, Norway}

\author{\O ystein Elgar\o y}
\email{Oystein.Elgaroy@astro.uio.no}
\affiliation{Institute of Theoretical Astrophysics , University of Oslo, Box 1029 Blindern, 0315 Oslo, Norway}

\begin{abstract}
We derive explicit and exact expressions for the physical velocity of a free particle comoving with the Hubble flow as measured by a static observer, and for the frequency shift of light emitted by a comoving source and received, again,  by a static observer. The expressions make it clear that an interpretation of the redshift as a kind of Doppler effect only makes sense when the distance between the observer and the source vanishes exactly.
\end{abstract}

\maketitle

\section{Introduction\label{sec:intro}}

Elementary textbooks often state Hubble's law both as a proportionality between the redshifts of galaxies and their distances, {\it and} as a proportionality between their recession velocities and their distances \cite{textbooks}. However, as students learn later on, the two statements are not equivalent~\cite{harrison}, and both are inaccurate.
Galaxies are not receding through space, but the distances between them increase with time. The redshift is not a Doppler effect, but the result of this stretching effect. The correct version of Hubble's law is
\begin{equation}
\dot{l} = H(t) l(t),
\label{eq:eq1}
\end{equation}
where $l(t)$ is the proper distance between the observer and the light-emitting source, $H(t) =
\dot{a}/a$ is the time-dependent Hubble parameter, $a(t)$ is the scale factor in the Friedman-Robertson-Walker line element, and dots denote differentiation with respect to cosmic time $t$.
The cosmological redshift is given by the ratio of the scale factor at the time of observation $t_{\rm o}$ and at the time of light emission $t_{\rm e}$:
\begin{equation}
1+z = \frac{a(t_{\rm o})}{a(t_{\rm e})}.
\label{eq:eq2}
\end{equation}
The time derivative of the proper distance does not correspond to a physical velocity, in particular it does not give the relative velocity of the light source and the observer. And, of course, the redshift is not given by the special relativistic formula with $\dot{l}$ playing the role of the relative velocity.

The notion of expanding space and the interpretation of the cosmological redshift as a result of it has been discussed by several
authors \cite{chodorowski1,chodorowski2,francis,cook} and is sometimes challenged, most recently by Bunn and Hogg \cite{bunnhogg}. They have argued
that the redshifts of distant galaxies in the expanding universe may be viewed as kinematical shifts due to suitably defined relative velocities,
and that this is the most natural interpretation.  Their notion of relative velocity is such that by definition
\begin{equation}
\sqrt{\frac{c+v_{\rm rel}}{c-v_{\rm rel}}}=\frac{a(t_{\rm o})}{a(t_{\rm e})}=1+z,
\label{eq:eq3}
\end{equation}
where $v_{\rm rel}$ is claimed to be the velocity of the source at the time of light emission relative to the observer at the present time.  More precisely, as shown by Narlikar \cite{narlikar}, the formula (\ref{eq:eq3}) appears if one parallel-transports the four-velocity of the source to the observer's location along the worldline of the photons, and then consider the difference between the parallel-transported velocity and the local velocity as the relative velocity of the observer and the source. An alternative result was similarly obtained by Chodorowski \cite{chodorowski3} who considered parallel-transport of the four-velocity along a different geodesic.

This goes to show that one can make the cosmological redshift look like a Doppler shift if one so wishes. The question is whether this is the most useful thing to do. As argued by Faraoni \cite{faraoni}, the set of observers on which the comoving coordinates are based, clearly provide a natural choice of observers because they alone see the cosmic microwave background as isotropic.  And for those observers, as Faraoni argues, the redshift is a gravitational effect, expressed most naturally by equation (\ref{eq:eq2}).

In this paper we will strengthen Faraoni's argument. We derive an exact and explicit equation for the cosmological redshift. From this expression it will be clear that for comoving observers the redshift is a pure expansion effect, and that it can be interpreted as a Doppler shift  only when the distance between the observer and the source vanishes.

\section{The connection of Hubble's law with physical velocity}

In comoving coordinates $\left(t,r,\theta,\phi\right)$ the spacetime geometry of an expanding, homogeneous and isotropic universe is described by the Friedmann-Lemaitre-Robertson-Walker (FLRW) line element
\beq
ds^2 = -c^2dt^2 + a^2(t)\left[dr^2 + R_0^2S_k^2\left(r/R_0\right)\left(d\theta^2 + \sin\theta d\phi^2\right)\right]\,,
\label{eq:eq4}
\eeq
where $a(t)$ is the scale factor, $R_0$ is the curvature radius of the three-space and
\beq
S_k\left(x\right) = \left\{ \begin{array}{lll}
\sin x\,,  &   k>0\,,\\
x\,,       &   k=0\,,\\
\sinh x\,, &   k<0\,.
\end{array}
\right.
\label{eq:eq5}
\eeq

Our purpose now is to obtain the physical velocity of a free particle comoving with the Hubble flow as measured relative to a static observer.
The static observer is defined as an observer who remains at rest at a constant physical distance $l_{\rm s}$ relative to the comoving observer
at the center $r=0$. The physical distance from the center to the static observer is $l_{\rm s}=a(t)\,r_{\rm s}(t)$, where
$r_{\rm s}(t)$ denotes the instantaneous radial coordinate position of the static observer. Hence, $l_{\rm s}$ must satisfy the equation
\beq
\dot{l}_{\rm s} = \dota\,r_{\rm s} + a(t)\,\dot{r}_{\rm s}=0\,,
\label{eq:eq6}
\eeq
from which we obtain the following coordinate velocity of the static observer:
\beq
\dot{r}_{\rm s} = -H(t)\,r_{\rm s}(t)\,.
\label{eq:eq7}
\eeq
Here the dots denote derivatives with respect to the coordinate time $t$ and $H(t)=\dota/a(t)$ is the Hubble parameter.
To construct an orthonormal tetrad representing the local laboratory frame of the static observer, we next consider the static
observer's four-velocity $\mathbf{u}_{\rm s}=\left(dx^\mu/d\tau_{\rm s}\right)\mathbf{e}_\mu$, where $\tau_{\rm s}$ denotes the proper time of the static observer. From the four-velocity identity $\mathbf{u}_{\rm s}\cdot\mathbf{u}_{\rm s}=-c^2$ and the metric given by~(\ref{eq:eq4}), we obtain
\beq
-c^2\left(\frac{dt}{d\tau_{\rm s}}\right)^2 + a^2(t)\left(\frac{dr}{d\tau_{\rm s}}\right)^2 = -c^2\,,
\label{eq:eq8}
\eeq
where we have used that $d\theta=d\phi=0$ because the static observer performs radial motion in the comoving coordinates. Eliminating
$dr/d\tau_{\rm s}$ from this equation by using the relation $dr/d\tau_{\rm s}=\dot{r}\left(dt/d\tau_{\rm s}\right)$ in combination with
Eq.~(\ref{eq:eq7}) then gives
\beq
\frac{dt}{d\tau_{\rm s}} = \left(c^2-\dota^2\dot{r}_{\rm s}^2\right)^{-1/2}c\,.
\label{eq:eq9}
\eeq
Since the static observer's timelike unit basis vector $\mathbf{e}_{\hat{\tau}}^{\rm s}$ is given by the observer's four-velocity
$\mathbf{u}_{\rm s}$, this yields the result
\beq
\mathbf{e}_{\hat{\tau}}^{\rm s} = \frac{dx^\mu}{d\tau_{\rm s}}\,\mathbf{e}_\mu = \left(\mathbf{e}_t
+\dot{r}_{\rm s}\,\mathbf{e}_r\right)\frac{dt}{d\tau_{\rm s}}
=\frac{c}{\sqrt{c^2-\dota^2r_{\rm s}^2}}\left(\mathbf{e}_t-H(t)\,r_{\rm s}\,\mathbf{e}_r\right)\,.
\label{eq:eq10}
\eeq
The static observer's spacelike unit basis vector pointing in the $r$-direction of the comoving coordinates,
$\mathbf{e}_{\hat{r}}^{\rm s}$, can be written in terms of the coordinate basis vectors as
$\mathbf{e}_{\hat{r}}^s=a_{\rm s}^t\mathbf{e}_t+a_{\rm s}^r\mathbf{e}_r$, where
$a_{\rm s}^t$ and $a_{\rm s}^r$ denote the components of the unit basis vector in the coordinate basis. This spacelike unit basis vector must
satisfy the two equations $\mathbf{e}_{\hat{r}}^{\rm s}\cdot\mathbf{e}_{\hat{\tau}}^{\rm s}=0$ and
$\mathbf{e}_{\hat{r}}^{\rm s}\cdot\mathbf{e}_{\hat{r}}^{\rm s}=1$. Using the result of Eq.~(\ref{eq:eq10}) and the definition
$\mathbf{e}_\mu\cdot\mathbf{e}_\nu=g_{\mu\nu}$, the corresponding algebraic equations may be solved to give
\beq
a_{\rm s}^t = -\frac{\dota r_{\rm s}}{c\sqrt{c^2-\dota^2r_{\rm s}^2}}\,,\quad a_{\rm s}^r
= \frac{c}{a(t)\sqrt{c^2-\dota^2r_{\rm s}^2}}\,.
\label{eq:eq11}
\eeq
The remaining two spacelike unit basis vectors can be chosen to be aligned with the $\theta$- and $\phi$-directions such that
$\mathbf{e}_{\hat{\theta}}^{\rm s}=r^{-1}\mathbf{e}_\theta$ and $\mathbf{e}_{\hat{\phi}}^{\rm s}=\left(r\sin\theta\right)^{-1}\mathbf{e}_\phi$.
One can easily check that the set of unit basis vectors
$\left\{\mathbf{e}_{\hat{\tau}}^{\rm s},\mathbf{e}_{\hat{r}}^{\rm s},
\mathbf{e}_{\hat{\theta}}^{\rm s},\mathbf{e}_{\hat{\phi}}^{\rm s}\right\}$ satisfy
the requirements for an orthonormal basis all along the observers world line, i.e.,
$\mathbf{e}_{\hat{\mu}}^{\rm s}\cdot\mathbf{e}_{\hat{\nu}}^{\rm s}=\eta_{\mu\nu}$, where $\eta_{\mu\nu}$ is the Minkowski metric.
This set of unit basis vectors defines a local system of coordinates
$\left(\tau_{\rm s},\hat{x}_{\rm s}^r,\hat{x}_{\rm s}^\theta,\hat{x}_{\rm s}^\phi\right)$ in the neighborhood of the static observer's world
line such that the three coordinate axes $\left(\hat{x}_{\rm s}^r,\hat{x}_{\rm s}^\theta,\hat{x}_{\rm s}^\phi\right)$ points along the directions
of the orthonormal basis vectors $\left\{\mathbf{e}_{\hat{r}}^{\rm s},\mathbf{e}_{\hat{\theta}}^{\rm s},\mathbf{e}_{\hat{\phi}}^{\rm s}\right\}$
at each point on the world line\cite{misner}. In this so-called `proper reference frame' of the static observer a
particle's \emph{physical} velocity corresponds directly to specifying the set of coordinate components
$\left\{d\hat{x}_{\rm s}^\mu/d\tau_{\rm s}\right\}$, in analogy with measurements of velocities within the special theory of relativity.

To calculate the physical velocity of a free particle following the Hubble flow as observed in the static observer's
proper reference frame, we next consider the four-velocity $\mathbf{u}_{\rm c}$ of this comoving particle. Firstly, in the coordinate
basis of the comoving coordinates we obtain
\beq
\mathbf{u}_{\rm c} = \frac{dx^\mu}{dt}\,\mathbf{e}_\mu = \mathbf{e}_t\,,
\label{eq:eq12}
\eeq
since the comoving particle has fixed spatial coordinates and the coordinate time $t$ in comoving coordinates equals the
particle's proper time. Secondly, in the local coordinates of the static observer's proper reference frame the same four-velocity
is calculated to be
\beq
\mathbf{u}_{\rm c} = \frac{dx_{\rm s}^{\hat{\mu}}}{dt}\,\mathbf{e}_{\hat{\mu}}^{\rm s}
=\left(\mathbf{e}_{\hat{\tau}}^{\rm s} + \frac{d\hat{x}_{\rm s}^r}{d\tau_{\rm s}}\,\mathbf{e}_{\hat{r}}^{\rm s}\right)\frac{d\tau_{\rm s}}{dt}
=\left[\left(\frac{dt}{d\tau_{\rm s}}+a_{\rm s}^t\frac{d\hat{x}_{\rm s}^r}{d\tau_{\rm s}}\right)\mathbf{e}_t
+\left(\frac{dt}{d\tau_{\rm s}}\,\dot{r}_{\rm s}+a_{\rm s}^r\frac{d\hat{x}_{\rm s}^r}{d\tau_{\rm s}}\right)
\mathbf{e}_r\right]\frac{d\tau_{\rm s}}{dt}\,.
\label{eq:eq13}
\eeq
The expressions in Eqs.~(\ref{eq:eq12}) and (\ref{eq:eq13}) may now be equated to give the two equations
\beq
\left(\frac{dt}{d\tau_{\rm s}}+a_{\rm s}^t\frac{d\hat{x}_{\rm s}^r}{d\tau_{\rm s}}\right)\frac{d\tau_{\rm s}}{dt}=1\,,
\label{eq:eq14}
\eeq
and
\beq
\left(\frac{dt}{d\tau_{\rm s}}\,\dot{r}_{\rm s}+a_{\rm s}^r\frac{d\hat{x}_{\rm s}^r}{d\tau_{\rm s}}\right)\frac{d\tau_{\rm s}}{dt}=0\,.
\label{eq:eq15}
\eeq
Finally, by solving Eq.~(\ref{eq:eq15}) for the velocity $d\hat{x}_{\rm s}^r/d\tau_{\rm s}$ and then substituting the results of
Eqs.~(\ref{eq:eq7}), (\ref{eq:eq9}) and (\ref{eq:eq11}) for $\dot{r}_{\rm s}$, $dt/d\tau_{\rm s}$ and $a_{\rm s}^r$, respectively,
we obtain the result
\beq
\frac{d\hat{x}_{\rm s}^r}{d\tau_{\rm s}} = \dota r_{\rm s}(t) = H(t)\,l_{\rm s}\,.
\label{eq:eq16}
\eeq
Note here that $l_{\rm s}$ both plays the role of the instantaneous physical distance between the center $r=0$ and the
location of the free particle following the Hubble flow as well as the physical distance between the center and the
location of the static observer measuring the velocity of the free particle. In other words, the velocity
$d\hat{x}_{\rm s}^r/d\tau_{\rm s}$ is the relative velocity between the free particle and the static observer \emph{at} the point
where the free particle passes by the observer. This velocity conforms with the usual point of view within general
relativity that the notion of relative velocity is a local one.

Thus, from the analysis presented above it should be clear that Hubble's law should not be interpreted as a formula
yielding the relative velocity between the (comoving) observer located at the center $r=0$ and a free particle located at a
certain comoving distance away at the radial coordinate $r$. The rate of change of the instantaneous physical distance $\dot{l}$
between a pair of comoving particles that enters Eq.~(\ref{eq:eq1}) is not a well-defined physical velocity because the
particles are in general located at different points in the curved spacetime. Rather, as Eq.~(\ref{eq:eq16}) shows, Hubble's law
corresponds to the physical velocity of a comoving particle as measured with respect to the locally orthonormal frame of a static
observer located at a constant physical distance $l_{\rm s}$ from the central observer.

Lastly, it should be mentioned that the formula given in Eq.~(\ref{eq:eq16}) is physically valid only in the region
$l_{\rm s}<c/H(t)$ for the following reason. The line element for a static observer can be obtained from Eq.~(\ref{eq:eq1})
by putting $d\theta=d\phi=0$ and then substituting for $dr^2$ the result of Eq.~(\ref{eq:eq7}),
$dr_{\rm s}^2=H^2(t)\,r_{\rm s}^2(t)\,dt^2$, which yields $ds^2=\left(-c^2+H^2(t)\,l_{\rm s}^2\right)dt^2$.
Since a physical particle with non-zero rest mass must follow a timelike world line for which $ds^2<0$, it is clear that
static observers cannot exist for physical distances $l_{\rm s}\geq c/H(t)$. That is, two observers separated by a
proper distance that equals or exceeds $c/H(t)$ cannot remain at rest relative to each other, but must move
apart. This fact ensures that galaxies never recede at superluminal physical speeds, in agreement with the fundamental principles of
general relativity. See \cite{crawford} for further clarification of this point.

\section{The frequency shift of light measured by static observers}

In this section we examine the frequency shift of light emitted in the radial direction by a comoving source and received
by a static observer.

Assume that a source located at the constant comoving coordinate position $r=r_{\rm e}$ emits light rays with frequency
$\omega_{\rm e}$ at comoving time $t=t_{\rm e}$. The energy of an emitted photon as measured in the laboratory frame of the
comoving source is then given by
\beq
\hbar\omega_{\rm e} = -\mathbf{p}\cdot\mathbf{u}_{\rm c}=-g_{\mu\nu}p^{\mu}\left(u_{\rm c}\right)^{\nu}\,,
\label{eq:eq17}
\eeq
where $\mathbf{p}$ is the four-momentum of the photon and $\mathbf{u}_{\rm c}$ is the four-velocity of the comoving source.
Obtaining the metric components from Eq.~(\ref{eq:eq1}) and the four-velocity $\mathbf{u}_{\rm c}$ from Eq.~(\ref{eq:eq12}),
this reduces to the relation
\beq
p^t(t_{\rm e}) = \hbar\omega_{\rm e}/c^2\,.
\label{eq:eq18}
\eeq
Assume further that the emitted photon is received by a static observer located at the instantaneous radial coordinate
position $r=r_{\rm s}$ corresponding to the time of observation $t_{\rm o}$. The energy of the received photon as measured
in the laboratory frame of the static observer can be written as
\beq
\hbar\omega_{\rm o} = -\mathbf{p}\cdot\mathbf{u}_{\rm s} = -g_{\mu\nu}p^{\mu}\left(u_{\rm s}\right)^{\nu}\,,
\label{eq:eq19}
\eeq
where $\omega_{\rm o}$ denotes the observed frequency and $\mathbf{u}_{\rm s}$ is the four-velocity of the static observer.
Obtaining the components of $\mathbf{u}_{\rm s}$ at $t=t_{\rm o}$ from Eq.~(\ref{eq:eq10}), this implies
\beq
\hbar\omega_{\rm o} = \frac{c^2p^t(t_{\rm o})+a^2(t_{\rm o})H(t_{\rm o})
\,r_{\rm s}p^r(t_{\rm o})}{\sqrt{c^2-\dot{a}^2(t_{\rm o})\,r_{\rm s}^2}}\,c\,.
\label{eq:eq20}
\eeq
The radial component of the photon's four-momentum may now be eliminated by noting that the four-momentum is a null vector
and therefore satisfies the condition $\mathbf{p}\cdot\mathbf{p}=0$. This immediately yields the relation $p^r=-cp^t/a(t)$.
Substituting this result in Eq.~(\ref{eq:eq20}), we thereby obtain
\beq
p^t(t_{\rm o}) = \sqrt{\frac{c+\dot{a}(t_{\rm o})\,r_{\rm s}}{c-\dot{a}(t_{\rm o})\,r_{\rm s}}}\,\frac{\hbar\omega_{\rm s}}{c^2}\,.
\label{eq:eq21}
\eeq

To proceed with our calculations, we now introduce the tensor
\beq
K_{\mu\nu}=a^2(t)\left[g_{\mu\nu}+\left(u_{\rm c}\right)_{\mu}\left(u_{\rm c}\right)_{\nu}\right]\,,
\label{eq:eq22}
\eeq
which may be shown to be a killing tensor \cite{carroll} satisfying $\nabla_{(\sigma}K_{\mu\nu)}=0$.
From this it follows that if $v^{\mu}$ is the four-velocity of a particle moving along a geodesic, the quantity
\beq
K^2 = K_{\mu\nu}v^{\mu}v^{\nu}=a^2(t)\left[v_{\mu}v^{\mu}+\left(\left(u_{\rm c}\right)_{\mu}v^{\mu}\right)^2\right]
\label{eq:eq23}
\eeq
is a constant along the geodesic. In the particular case where the particle is a massless photon, we have $v^{\mu}\equiv p^{\mu}$
and $p_{\mu}p^{\mu}=0$. Eq.~(\ref{eq:eq23}) then reduces to
\beq
K = \sqrt{a^2(t)\left[g_{\mu\nu}\left(u_{\rm c}\right)^{\mu}p^{\mu}\right]^2} = c^2a(t)p^t = \mbox{const.}\,,
\label{eq:eq24}
\eeq
since the only non-vanishing component of the comoving observer's four-velocity is $\left(u_{\rm c}\right)^{t}=1$ (Eq.~(\ref{eq:eq12})).
Because the quantity $c^2a(t)p^t$ is a constant along the trajectory of the emitted photon, it now follows that
\beq
c^2a(t_{\rm e})\,p^t(t_{\rm e}) = c^2a(t_{\rm o})\,p^t(t_{\rm o})\,,
\label{eq:eq25}
\eeq
or, utilizing Eqs.~(\ref{eq:eq18}) and (\ref{eq:eq21}),
\beq
\frac{\omega_{\rm o}}{\omega_{\rm e}}
= \sqrt{\frac{c-\dot{a}(t_{\rm o})\,r_{\rm s}}{c+\dot{a}(t_{\rm o})\,r_{\rm s}}}\,\frac{a(t_{\rm e})}{a(t_{\rm o})}\,.
\label{eq:eq26}
\eeq
Finally, by inserting the result of Eq.~(\ref{eq:eq16}) in~(\ref{eq:eq26}), the relation between the emitted and observed frequencies
can be rewritten as
\beq
\frac{\omega_{\rm o}}{\omega_{\rm e}}
= \sqrt{\frac{1-\left(d\hat{x}_{\rm s}^r/d\tau_{\rm s}\right)/c}
{1+\left(d\hat{x}_{\rm s}^r/d\tau_{\rm s}\right)/c}}\,\frac{a(t_{\rm e})}{a(t_{\rm o})}\,.
\label{eq:eq27}
\eeq

The expression in Eq.~(\ref{eq:eq27}) reveals that the relation between the emitted and observed frequencies in general is a product
of the conventional special-relativistic Doppler shift factor,
\beq
\sqrt{\frac{1-\left(d\hat{x}_{\rm s}^r/d\tau_{\rm s}\right)/c}{1+\left(d\hat{x}_{\rm s}^r/d\tau_{\rm s}\right)/c}}\,,
\label{eq:eq28}
\eeq
and the well-known gravitational redshift factor,
\beq
\frac{a(t_{\rm e})}{a(t_{\rm o})}\,.
\label{eq:eq29}
\eeq
Two limiting cases are of particular interest here. Firstly, if the radial coordinate position of the comoving source instantaneously
coincides with the radial coordinate position of the static observer at the time of emission $t_{\rm e}$, then we have
$t_{\rm e}=t_{\rm o}$ and therefore $a(t_{\rm e})/a(t_{\rm o})=1$. The general expression given by Eq.~(\ref{eq:eq27}) thereby simplifies
to the expression in~(\ref{eq:eq28}). Thus, in this case, the shift in frequency can indeed be ascribed to a pure Doppler effect
caused by the relative velocity between the observer and emitter. This result can be understood by noting that, in the immediate
neighborhood of the origin of the local reference frame associated with the static observer, it is impossible to detect any effects of
spacetime curvature. Accordingly, the static observer will interpret the observed frequency shift as having a special-relativistic
kinematical origin and not as an effect caused by the curvature of spacetime.

Secondly, if the static observer is located at the origin, i.e., $r_{\rm s}=0$, then we have
$d\hat{x}_{\rm s}^r/d\tau_{\rm s}=\dot{a}(t_{\rm o})\,r_{\rm s}=0$ and the general expression given in~(\ref{eq:eq27}) thereby reduces
to the expression in~(\ref{eq:eq29}). Note that the static observer is now comoving. Hence, in this case, the Doppler effect vanishes
\emph{identically} and the exact formula for the frequency shift takes the form of a pure gravitational redshift for \emph{all} distances
between emitter and observer. We emphasize that this result is exact and therefore remains true even if the distance between emitter
and observer is much smaller than the Hubble radius $c/H(t_{\rm o})$. As a consequence the redshift observed by comoving observers may in
general be attributed to the effect of an expanding space. The Doppler shift is relevant only to observers having a non-vanishing peculiar
velocity $\dot{r}$.

\section{Summary and conclusion}

We have derived formulae for the physical velocity and redshift of comoving objects as measured by
a static observer in an expanding, homogenous and isotropic universe. These expressions make it clear that for comoving observers the cosmological redshift is a gravitational effect, determined by the scale factor $a(t)$ in all cases except when the distance between the observer and the source vanishes exactly. Nothing prevents one from {\it defining} a velocity to make the cosmological redshift look like a Doppler shift, but the velocity thus defined is unphysical. For observers like us, the comoving coordinates of the FLRW line element are the most natural choice, and in those coordinates the cosmological redshift is best interpreted as a result of the increase of $a(t)$ with cosmic time, or, to put it shortly, the expansion of space.

\end{document}